\documentclass[useAMS,usenatbib]{mn2e}

\usepackage{graphicx}

\title[Bright spectrophotometric standards]{New bright optical 
spectrophotometric standards: A-type stars from the STIS Next 
Generation Spectral Library}
\author[C. Allende Prieto and C. del Burgo]{C. Allende Prieto$^{1,2}$\thanks{E-mail:
callende@iac.es} and C. 
del Burgo$^{3}$\\
$^{1}$Instituto de Astrof\'{\i}sica de Canarias, 38205 La Laguna, 
Tenerife, Spain\\
$^{2}$Departamento de Astrof\'{\i}sica, Universidad de La Laguna, 
38206 La Laguna, Tenerife, Spain \\
$^{3}$Instituto Nacional de Astrof\'{\i}sica, \'Optica y Electr\'onica, 
Luis Enrique Erro 1, Sta. Ma. Tonantzintla, 72840 Puebla, Mexico}

\begin{document}

\date{}

\pagerange{\pageref{firstpage}--\pageref{lastpage}} \pubyear{2002}

\maketitle

\label{firstpage}

\begin{abstract}
Exoplanets have sparked interest in extremely high signal-to-noise ratio
spectroscopic observations of very bright stars, in a regime where flux calibrators, 
 in particular DA white dwarfs, are not available. We argue that A-type
stars offer a useful alternative and reliable space-based 
spectrophotometry is now available for
a number of bright ones in the range $3<V<8$ mag. By means of comparing
observed spectrophotometry and model fluxes, we identify 18 new  
very-bright trustworthy A-type flux standards for the optical  range (400-800 nm),
 and provide scaled model fluxes for them. 
Our tests suggest that the absolute fluxes for these stars in the optical 
are reliable to within 3\%.  We limit the spectral range to
400-800 nm, since our models have difficulties to reproduce
the observed fluxes in the near-infrared and, especially, in the near-UV, 
where the discrepancies rise up to $\sim$ 10\%. Based on our model fits,
we derive angular diameters with an estimated accuracy of about 1\%.

\end{abstract}


\section{Introduction}

Absolute flux calibration of astronomical spectra has always proven
challenging, in particular from the ground, but it is often crucial 
for understanding the nature of astronomical sources. 
Progress on this field has been summarized in the papers
presented in two dedicated conferences (Sterken 2007; 
Milone \& Sterken 2011)  and in a recent review  
by Bohlin et al. (2014).

Reliable standard stars are needed at all wavelengths. In the optical, 
 efforts based on DA white dwarf modeling for HST calibration have
paid off, and good consistency is found around the system based on
three primary stars, and a longer list of secondary ones 
(Bohlin 2007, 2010, Bohlin \& Cohen 2008, Bohlin \& Koester 2008, 
and references therein).
Those secondary stars span a range in Johnson $V$ approximately between $-1$ 
and 17, but there are few calibrators available on both ends, 
at $V$ magnitudes brighter than 6 or fainter than 15.
Cosmology surveys are now pushing the faint limit, and exoplanet research 
is focusing on stars significantly brighther than tenth magnitude.

In a previous paper, we contributed a number of faint 
DA white dwarf standards identified from the 
Sloan Digital Sky Survey (Allende Prieto, Hubeny \& Smith 2009). 
In this work we aim at covering the difficult area of the 
bright sources, at $V<8$ mag. No white dwarfs are known in this range, 
so other sources must be considered. 

While any spectral type may be useful at very low resolution,
standard calibrators usable over a wide range in spectral resolution, 
should have a smooth continuum. As for DA white dwarfs, being able to 
compute accurately the fluxes for a calibrator implies that its nature
is well understood, discarding the presence of composite sources, 
significant interstellar extinction, or systematic observational errors. 
A-type stars
seem the most natural kind of calibrators that are available to us in the
range $0<V<10$ mag.

A-type stars are warm enough that their spectra are relatively clean of
metal lines in the optical range, although this is not true in the UV. 
A-type stars have fully radiative envelopes, avoiding the problem of
convective energy transport, which is very hard to model accurately
and can have an important influence on the thermal structure of 
 deep photosphere where the continuum and the wings of  
Balmer lines form. With the shape of the continuum in the optical 
being dominated by  photoionization of hydrogen from the $n=2$ level, 
modeling the spectral energy distribution of these stars in the 
visible is fairly  independent of the detailed abundances of  
metals such as magnesium, silicon, or iron, which 
are important electron donors. Empirically, an excellent
agreement has been found between observed and model spectral energy distributions
of A-type stars in the 0.8 to 2.5 $\mu$m range by Bohlin \& Cohen (2008).

On a negative note, A-type stars, unlike cooler dwarfs, 
tend to be young and rotate at a high rate. Fast rotation can drive the 
shape of the star away from a sphere, significantly altering 
the photospheric effective temperature and effective gravity 
across the surface of the star. For extreme cases,
rotation can modify profoundly the spectral energy distribution or
the photometric colors of a star (P\'erez Hern\'andez et al. 1999).
Fortunately, as discussed by Gulliver, Hill \& Adelman 1994  
for the case of Vega (see also 
Aufdenberg et al. 2006 and Hill, Gulliver \& Adelman 2010), 
despite there are spectral windows that 
are  susceptible to rotation, the optical happens to 
be fairly insensitive to this parameter.
It has also been found that a high  
fraction of A-type stars, in particular at young ages, 
have dusty circumstellar disks (e.g. Su et al. 2006, 
Currie, Plavchan \& Kenyon 2008), which may affect their
energy distributions in the infrared but are innocuous in the 
optical. Very fast rotators can have gaseous disks, but these are not
expected to alter the observed spectral energy distribution in the
optical (Abt, Tan \& Zhou 1997). In summary, A-type stars can be
suitable flux standards at visible wavelenghts, and are 
particularly useful to fill 
the gap left by DA white dwarfs at bright magnitudes. 

The STIS Next Generation Spectral Library (NGSL; Gregg et al. 2006)
includes spectrophotometry for 370 stars of a wide range of
spectral types, including about 70 A-type stars with Johnson $V$
magnitudes spanning 2.9--10.3 mag. 
We have examined closely the A-type stars in this
library and compared them with calculated fluxes based on 
Kurucz ODFNEW model atmospheres (Castelli \& Kurucz 2004). Taking 
good agreement between  model fluxes and observations 
as indicative that the star is a {\it normal} A-type star and
that the observations do not suffer from significant instrumental
distortions, we have selected a number of stars that appear to be 
good candidates for flux standards. 
Similar to what has been done with DA white dwarfs (see, e.g., Bohlin 2007), 
model fluxes  are scaled by reliable broad-band photometry to 
set the absolute scale, and if no photometric variability is found, 
the stars are promoted  to spectrophotometric standards.

Section \ref{basic} describes the data and models used in this study. 
\S \ref{ferre} is devoted to the analysis of  
observed spectral energy distributions (SEDs), and 
defines our criteria to select reliable standards based on 
the goodness-of-fit and the photometric stability
of the sources.  \S \ref{absolute}
focuses on the available photometry and the appropriate absolute
scale for the stellar fluxes, while \S \ref{diameter} provides
inferred angular diameters for our sample, which are compared to interferometric
determinations available for a few objects. 
The paper closes with a short summary and our conclusions.

\section{Observed and model spectra}
\label{basic}

The two main ingredients involved in our analysis are the STIS
spectrophotometry from the NGSL, and  synthetic spectra based on 
model atmospheres. We 
discuss below in \S \ref{absolute} other observations used to check 
the absolute  zero point to be adopted for the standard calibrators.

\subsection{Spectrophotometric observations}
\label{spectra}


Photometric and spectrophotometric observations from space benefit 
from a dramatic reduction in atmospheric extinction 
and  variability compared to data taken from the ground. 
In this spirit, an HST snapshot programme 
(GO 9088, 786, 10222) was granted observing time to obtain 
STIS spectrophotometry for 374 stars between 
$\sim$ 0.2 and 1.0 $\mu$m in cycles 10, 12, and 13
(Gregg et al. 2006). Several other successful GO and AR proposals
aimed at improving the corrections for the slit throughput and other
issues followed. The first version of the library was made
public in 2008 and, with  additional corrections,
the second version was released in March 2010. Readers are referred 
to the online documentation provided with the library for details
(Heap \& Lindler 2007; Lindler \& Heap 2008\footnote{http://archive.stsci.edu/pub/hlsp/stisngsl/aaareadme.pdf}).

The data are fairly homogeneous, even though the wide spectral coverage
required a combination of  three different instrumental setups with three
different gratings (G230LB, G430L and G750L). 
The expected resolving power $R$\footnote{$R\equiv \lambda/\delta\lambda$, 
where $\delta\lambda$ corresponds
to the FWHM of a Gaussian line spread function} 
varies slightly with wavelength, but 
is nominally about $500 < R < 1100$, as
described in the STIS Instrument 
Handbook\footnote{http://www.stsci.edu/hst/stis/documents/\\
/handbooks/currentIHB/cover.html}.

The spectra were subject to a systematic analysis based on model
spectra by Castelli \& Kurucz (2004) to derive atmospheric parameters,
constrained by the parallaxes measured by the Hipparcos mission
(van Leeuwen 2007) and stellar evolution models by
Vandenberg et al. (2006). To some extent the analysis performed
in \S \ref{ferre} duplicates this work, but we chose not to use
parallaxes and stellar evolution models,  given that 
we are mainly concerned with the models 
providing an accurate description of the stars' spectral
energy distributions, and not so much unbiased atmospheric
parameters. In fact, metallicity and surface gravity have 
a very limited impact on the spectral energy distribution of 
these stars in the wavelength range 400-800 nm.
For example, the relative (normalized by their average value) 
fluxes of an A0V star like Vega change  by less than half a
percent when the surface gravity and metallicity vary by as much 
as 0.5 dex (del Burgo, Allende Prieto \& Peacocke 2010).
Independent estimates of the atmospheric parameters for NGSL
stars have been provided by  
Koleva \& Vazdekis (2012), with whom we compare results.

\subsection{Calculated spectral fluxes}
\label{synthesis}


We have computed a grid of synthetic spectra covering the
wavelength range 200--1000 nm with wavelength steps 
equivalent to 0.6 km s$^{-1}$. The calculations are based on
Kurucz ODFNEW model atmospheres (Castelli \& Kurucz 2004) 
and the synthesis code ASS$\epsilon$T 
(Koesterke 2009; Koesterke, Allende Prieto \& Lambert 2008), 
operated in 1D (plane-parallel geometry) mode. 
The reference solar abundances for the synthesis are from 
Asplund, Grevesse \& Sauval (2005), and these were scaled for
elements heavier than helium according to the chosen  metallicities. 
The micro-turbulence was set to 2 km s$^{-1}$.
While temperature and density are taken from the model
atmospheres, the electron density is recalculated for consistency
with the equation of state used, which includes the first 92 elements
in the period table and 338 molecules (Tsuji 1964, 1973, 1976, with some
updates). Partition functions are adopted from  Irwin (1981).

Bound-free absorption from H, H$^-$, HeI, HeII,  and 
the first two ionization stages of C, N, O, Na, Mg,  Al, Si, 
Ca (from the Opacity Project; see Cunto et al. 1993) 
and Fe (from the Iron Project; Bautista 1997; Nahar 1995) are included.
Line absorption is included in detail from the atomic and molecular
(H$_2$, CH, C$_2$, CN, CO, NH, OH, MgH, SiH, and SiO) files 
compiled by Kurucz\footnote{kurucz.harvard.edu}, but molecules
are not important at the temperatures we are interested in. 
Level dissolution near the Balmer series limit is accounted
for (Hubeny et al. 1994).
The radiative transfer calculations include  
 Rayleigh (H; Lee \& Kim 2004) 
and electron (Thomson) scattering. The damping of H lines
is treated in detail using Stark (Stehl\'e 1994; 
Stehl\'e \& Hutcheon 1999) and self
broadening (Barklem, Piskunov \& O'Mara 2000 
for Balmer and Ali \& Griem 1966  for Lyman and Paschen lines). 
We used a grid with values of effective temperature in the range
$7500 \leq T_{\rm eff} \leq 13,000$ K,
surface gravity 
$2.5 \leq \log g \leq 5.0$, and  metallicity 
$-2.5 \leq$[Fe/H]$\leq +0.5$, with
steps of 250 K, 0.5 dex, and 0.5 dex, respectively. 
Spectra were convolved with a Gaussian kernel to
account for instrumental broadening.

\section{Spectral analysis}
\label{ferre}


We searched for the values of $T_{\rm eff}$, $\log g$,  [Fe/H], and resolving 
power
associated with the model fluxes that were closest to the
NGSL spectra in the 200-1000 nm wavelength range for selected stars. 
A straight $\chi^2$ was used as merit function, making no difference
between lines and continuum regions.
Interstellar reddening was neglected, and we assumed that the correction of 
the Doppler velocities to the rest frame in the library is perfect, although we noticed
slight inconsistencies in the wavelength scale of the spectra in the blue.
The optimization was
done using the FORTRAN90 code {\tt FER\reflectbox{R}E}\footnote{Available
from http://hebe.as.utexas.edu/ferre} 
(Allende Prieto 2004, Allende Prieto et al. 2006, 2008, 2009), 
using  the Nelder-Mead algorithm (Nelder \& Mead 1965). Model
fluxes with any set of atmospheric parameters are derived by
cubic B\'ezier interpolation (see Auer 2003 and included references) 
in the grid of model fluxes described in \S \ref{synthesis}.

\subsection{The prime A-type standard: Vega}
\label{vega}


Vega offers an excellent opportunity to test our analysis
procedure on a reference star that is actually considered
an excellent standard. The Vega spectrum obtained with
STIS  and calibrated on the flux
scale defined by  three white dwarfs that constitute the set of 
primary HST spectrophotometric standards was found in excellent
agreement with the fluxes predicted by Kurucz models with
parameters $T_{\rm eff} \simeq 9550$ K, $\log g = 3.95$, and 
[Fe/H] $=-0.5$ (Bohlin \& Gilliland 2004; see also 
Garc\'{\i}a-Gil et al. 2005).  More recently, after
applying corrections for charge-transfer inefficiency in the
STIS CCDs, the HST fluxes for this star have changed
by up to 2\%, leading to a lower temperature estimate of about 
$T_{\rm eff} \simeq 9400$ K (Bohlin 2007). 

\begin{figure}
\centering
\includegraphics[height=58mm,angle=0]{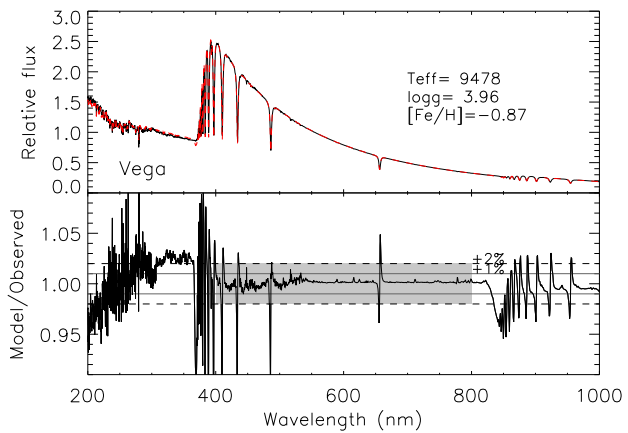}
 \caption{Upper panel: Observed (solid black line) and best-fitting model (broken red line) for 
 Vega. Bottom panel: Ratio between model and observed fluxes.}
\label{f1}
\end{figure}

We analyze the revised STIS spectrophotometry for Vega exactly
in the same way as the NGSL A-type stars.
We obtain $T_{\rm eff} = 9485 \pm 10$ K, $\log g = 3.94 \pm 0.01$, 
[Fe/H]$=-0.85 \pm 0.01$,
and a resolving power $R=700 \pm 5$, with good agreement between 
 model and observed fluxes,
as ilustrated in Fig. 1, and a reduced chi-squared value of 3.6. 
These results are overall consistent with del Burgo et al. (2010).
We adopted systematic and random error in the fluxes as provided by Bohlin \& Gilliland, 
where the former is 1\% and dominates at the wavelengths used in our analysis.
Our quoted uncertainties include random and systematic ones.
Random uncertainties are derived either from the inversion of the curvature matrix,
or from Monte Carlo experiments adding Gaussian noise to the spectrum, with very
similar results, and 
we estimate systematic uncertainties by reanalyzing the spectrum after 
adding the maximum systematic error to the fluxes, with different sign at opposite 
extremes of our wavelength range, and intermediate values inbetween.
Bohlin (2007) finds that the resolving power is $R \simeq 500$, slightly lower 
than our preferred value.

\begin{table*}
        \centering
        \caption{Derived parameters and angular diameters ($\theta$) for the proposed bright spectrophotometric
	  A-type standards.  The {\it scatter} given corresponds to the standard deviation in   
	  $\theta$ for the spectral data points in the 400-800 nm range. The uncertainty $\sigma(\theta)$ 
	  is derived by adding the scatter in quadrature with a 0.4\% error that corresponds to propagating 
	  the uncertainty in the zero-point of the flux scale (0.8 \% or 0.008 magnitudes). Both the
	  atmospheric parameters from the NGSL library (Lindler \& Heap 2008) and those derived in this 
	  paper are provided. We estimate our uncertainties are about
	  120 K, 0.2 dex, and 0.1 dex (random), and 
	  150 K, 0.2 dex, and 0.1 dex (systematic), for 
	  $T_{\rm eff}$, $\log g$, and [Fe/H], respectively (but see text for a caveats about $\log g$
	  for some stars).
	  The $V$-band photometry derived from the NGSL spectra, and those measured
	  from the ground and compiled by Mermilliod et al. (1997) are in the last two columns.}
        \label{tab:example_table}
        \begin{tabular}{lrrrrrrrrrrr} %
                \hline
                      &           &         &                    &
                      \multicolumn{3}{|c|}{--------- This work ---------} & 
                      \multicolumn{3}{|c|}{------------ NGSL ------------} & & \\
                Star  &  $\theta$ & scatter & $\sigma(\theta)$  &
                 $T_{\rm eff}$  & $\log g$  & [Fe/H]  &  
                $T_{\rm eff}$ & $\log g$  & [Fe/H]    & 
                V (NGSL)  &  V (phot.) \\
                      &    (mas)    &     (mas) &      (mas)        & 
                      (K)         &    (cm s$^{-2}$) &     &  
                      (K)         &    (cm s$^{-2}$) &      \\
                \hline
 HD   319 &  0.2786 &  0.0014 &  0.0018 &   8054 &  3.55 & -0.60 &   8195 &  3.86 & -0.39 & 	5.9156 &  5.932 $\pm 0.004$ \\
 HD 18769 &  0.2760 &  0.0020 &  0.0023 &   8057 &  3.63 &  0.15 &   8381 &  4.13 &  0.49 & 	5.8980 & 5.905 $\pm 0.014$\\
 HD 34797 &  0.1242 &  0.0007 &  0.0009 &  12619 &  4.00 & -0.19 &  12884 &  4.06 &  0.02 & 	6.5283 & 6.532 $\pm 0.040$\\
 HD 38237 &  0.1613 &  0.0010 &  0.0012 &   7901 &  3.60 & -0.07 &   8100 &  3.91 &  0.07 & 	7.1495 & \dots \\
 HD 40573 &  0.1004 &  0.0007 &  0.0008 &  10065 &  4.21 & -0.30 &  10200 &  4.20 & -0.37 & 	7.4679 & \dots \\
 HD 78316 &  0.2222 &  0.0011 &  0.0014 &  12747 &  3.79 & -0.26 &  12442 &  3.71 & -0.12 & 	5.2410 & 5.233 $\pm 0.008$\\
 HD 79469 &  0.5027 &  0.0027 &  0.0034 &  10453 &  4.17 & -0.22 &  10489 &  4.23 & -0.18 & 	3.8702 & 3.885 $\pm 0.009$\\
 HD 97633 &  0.7566 &  0.0036 &  0.0047 &   9136 &  3.50 &  0.01 &   9107 &  3.58 & -0.17 & 	3.3031 & 3.338 $\pm 0.023$\\
 HD 110073 &  0.3078 &  0.0024 &  0.0027 &  12086 &  3.81 & -0.78 &  12041 &  3.77 & -0.37 & 	4.6605 & 4.638 $\pm 0.004$\\
 HD 141795 &  0.7559 &  0.0057 &  0.0065 &   8094 &  3.76 &  0.07 &   8418 &  4.21 &  0.34 & 	3.6976 & 3.708 $\pm 0.009$ \\
 HD 141851 &  0.3975 &  0.0020 &  0.0026 &   8135 &  3.77 & -0.31 &   8231 &  3.96 & -0.22 & 	5.1075 & 5.105 $\pm 0.005$\\
 HD 164967 &  0.1529 &  0.0012 &  0.0013 &   8153 &  3.84 & -0.75 &   8533 &  4.07 & -0.60 & 	7.1957 & \dots \\
 HD 166991 &  0.1730 &  0.0008 &  0.0011 &   8347 &  3.86 & -0.29 &   8496 &  4.00 & -0.26 & 	6.8107 & \dots \\
 HD 174240 &  0.2061 &  0.0013 &  0.0016 &   8894 &  3.69 & -0.26 &   9274 &  3.76 & -0.18 & 	6.2177 & 6.240 $\pm$\dots \\
 HD 196426 &  0.1484 &  0.0007 &  0.0009 &  12376 &  3.85 & -0.63 &  12800 &  3.95 & -0.48 & 	6.2023 & 6.211 $\pm 0.015$\\
 HD 201377 &  0.2020 &  0.0012 &  0.0014 &   7905 &  3.59 & -0.10 &   8084 &  3.92 &  0.04 & 	6.6725 & \dots \\
 HD 204041 &  0.2254 &  0.0011 &  0.0014 &   7999 &  3.86 & -0.90 &   8259 &  4.17 & -0.60 & 	6.4354 & 6.455 $\pm 0.005$\\
 HD 205811 &  0.1977 &  0.0009 &  0.0012 &   9294 &  4.12 & -0.00 &   9286 &  4.20 & -0.13 & 	6.1654 & 6.193 $\pm 0.011$\\
        \end{tabular}
\end{table*}

The average ratio $<F/f>$ between the flux
predicted by the model at the stellar surface and that observed
in the 400-800 nm spectral window is then used to derive a stellar 
angular diameter of 
$\theta \simeq 2 R/d = 2 \sqrt{f/F} = 3.333 \pm 0.013$ mas.
The random error contribution estimated from the scatter of the flux
ratio across the selected wavelength interval is about 0.0004 mas, much smaller
than the 0.8\% uncertainty in the absolute magnitude in the $V$ band 
for Vega (Bohlin 2007), which sets the absolute flux calibration of the spectrum
and  dominates the error budget.
This value is very similar to that inferred in the same fashion 
by Bohlin (2007; $\theta= 3.335$ mas if $T_{\rm eff}=9400$ K but
$\theta=3.273$ mas if $T_{\rm eff}=9550$ K), and to the results of
Aufdenberg et al. (2006), $\theta= 3.33 \pm 0.01$ mas using a model including
distortions due to fast rotation. Monnier et al. et al. (2012) arrived at 
$\theta= 3.324$ mas  based on interferometric measurements, but see 
Peterson et al. (2006) for a discrepant result.  

This excellent agreement gives
us confidence that the analysis procedure and the assumptions 
involved are valid, but it should be noted  that Vega is a rapidly 
rotating pole-on star and it may be somewhat variable -- see the
discussion of these and other problems with Vega by Gray (2007) and also Bohlin (2014) 
-- limiting the extend to which the simple hydrostatic 
plane-parallel models used here can reproduce its spectral energy distribution. 
Note that the predicted fluxes are systematically higher 
in the region between 280 and
370 nm, in line with the expected corrections for  rotationally distorted
models (Gulliver, Hill \& Adelman 1994).

\subsection{The bright A-type stars in NGSL}
\label{stars}

\begin{figure*}
\centering
\includegraphics[height=58mm,angle=0]{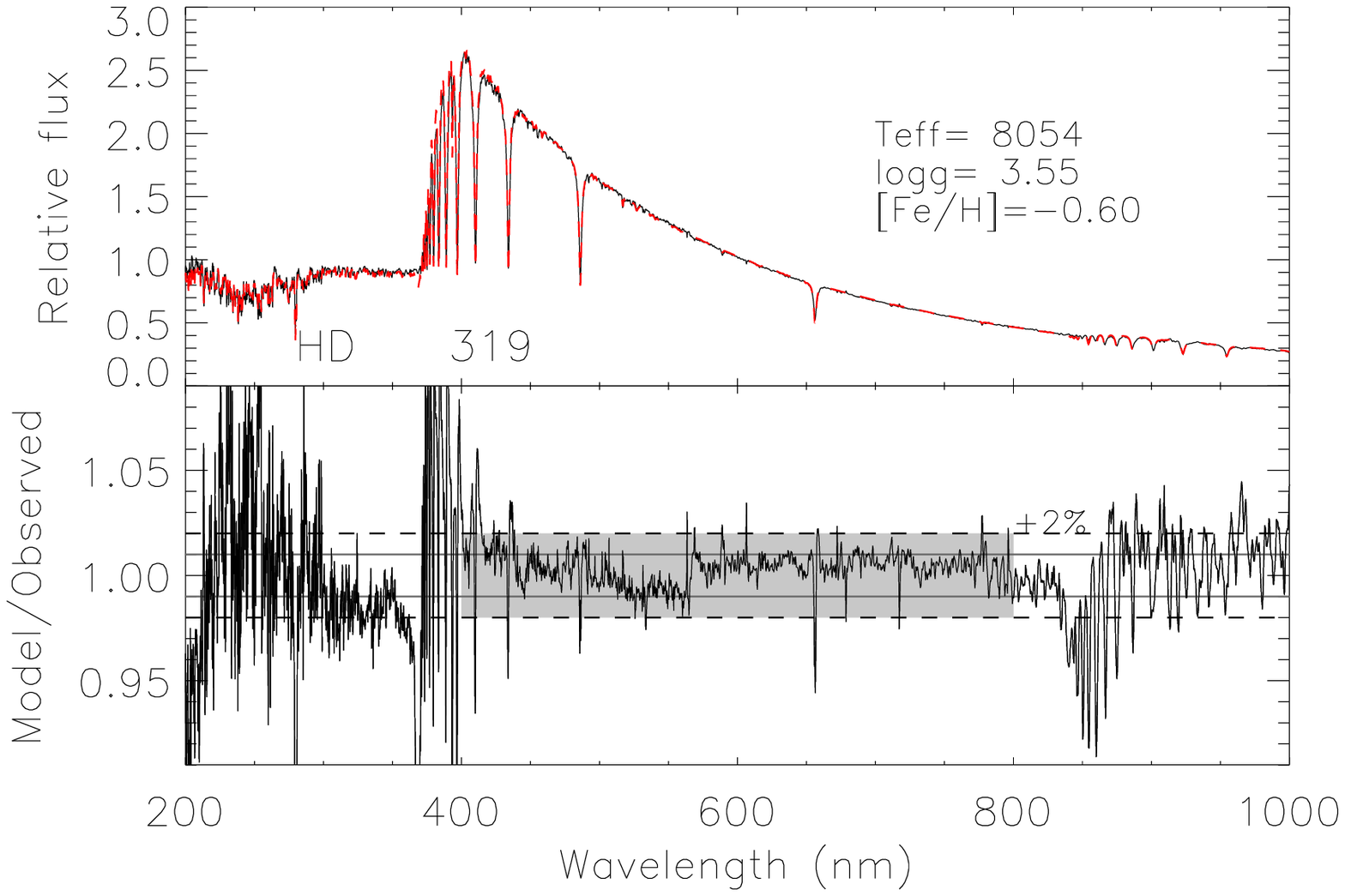}
\includegraphics[height=58mm,angle=0]{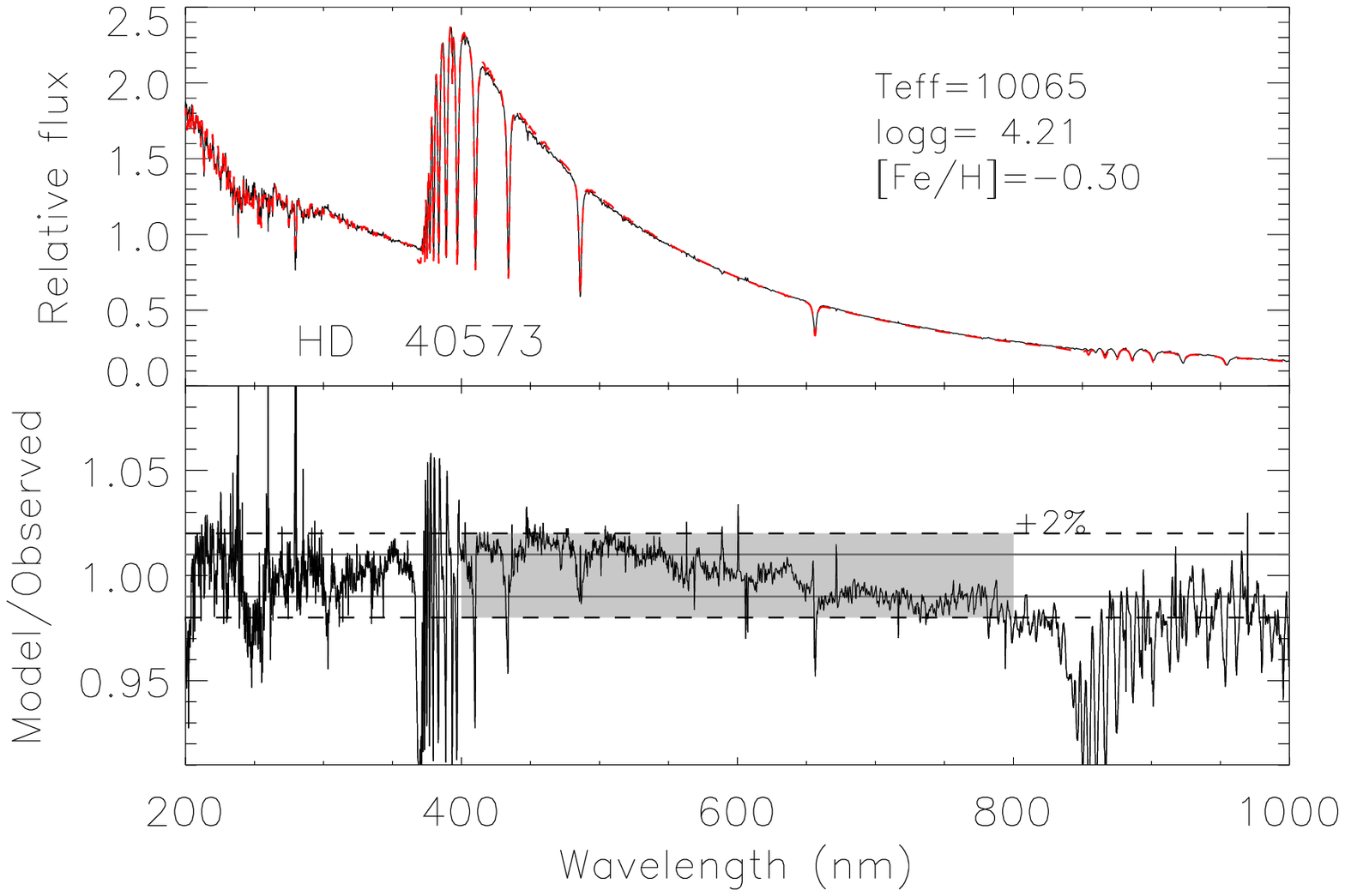}
\includegraphics[height=58mm,angle=0]{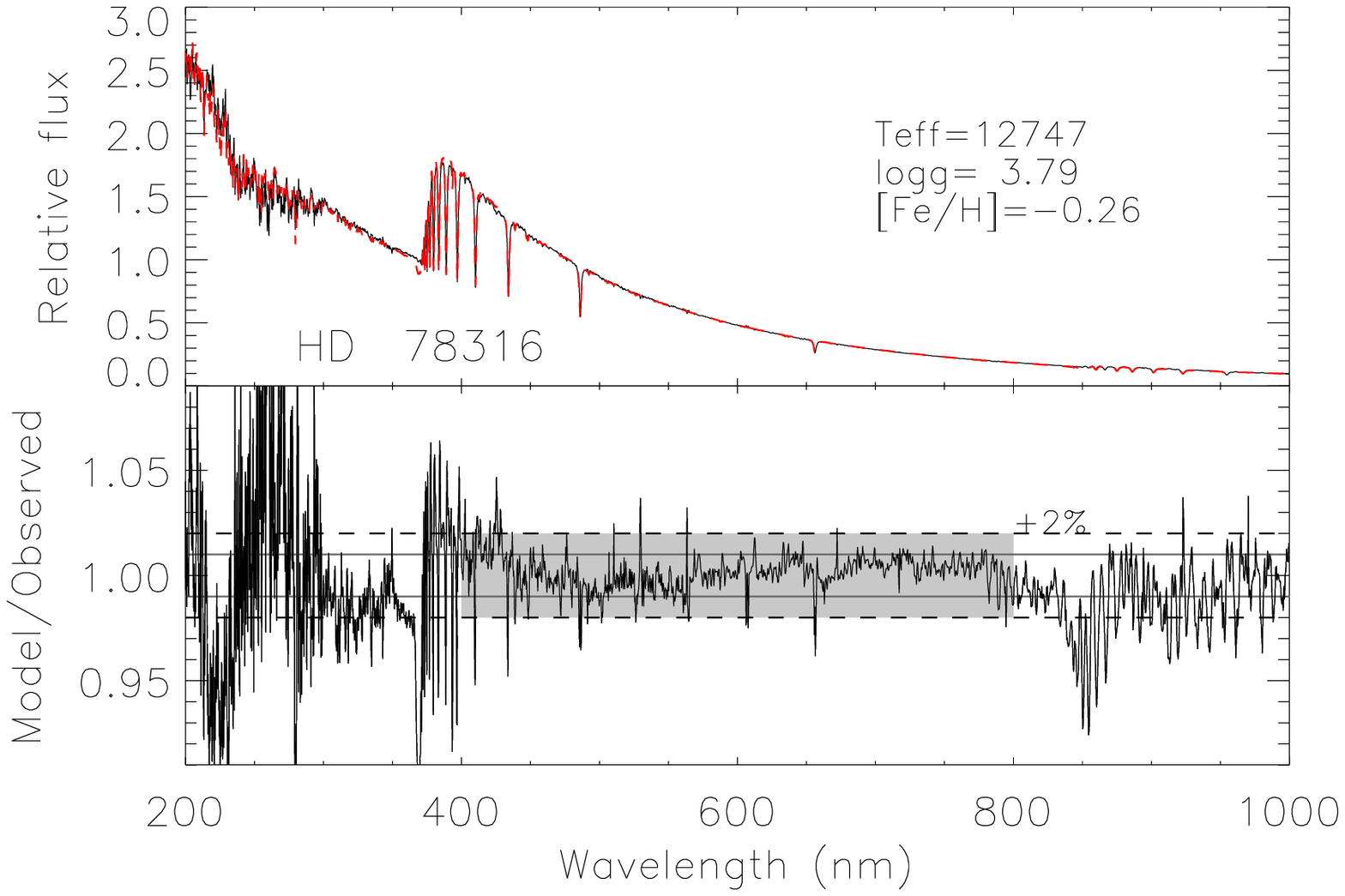}
\includegraphics[height=58mm,angle=0]{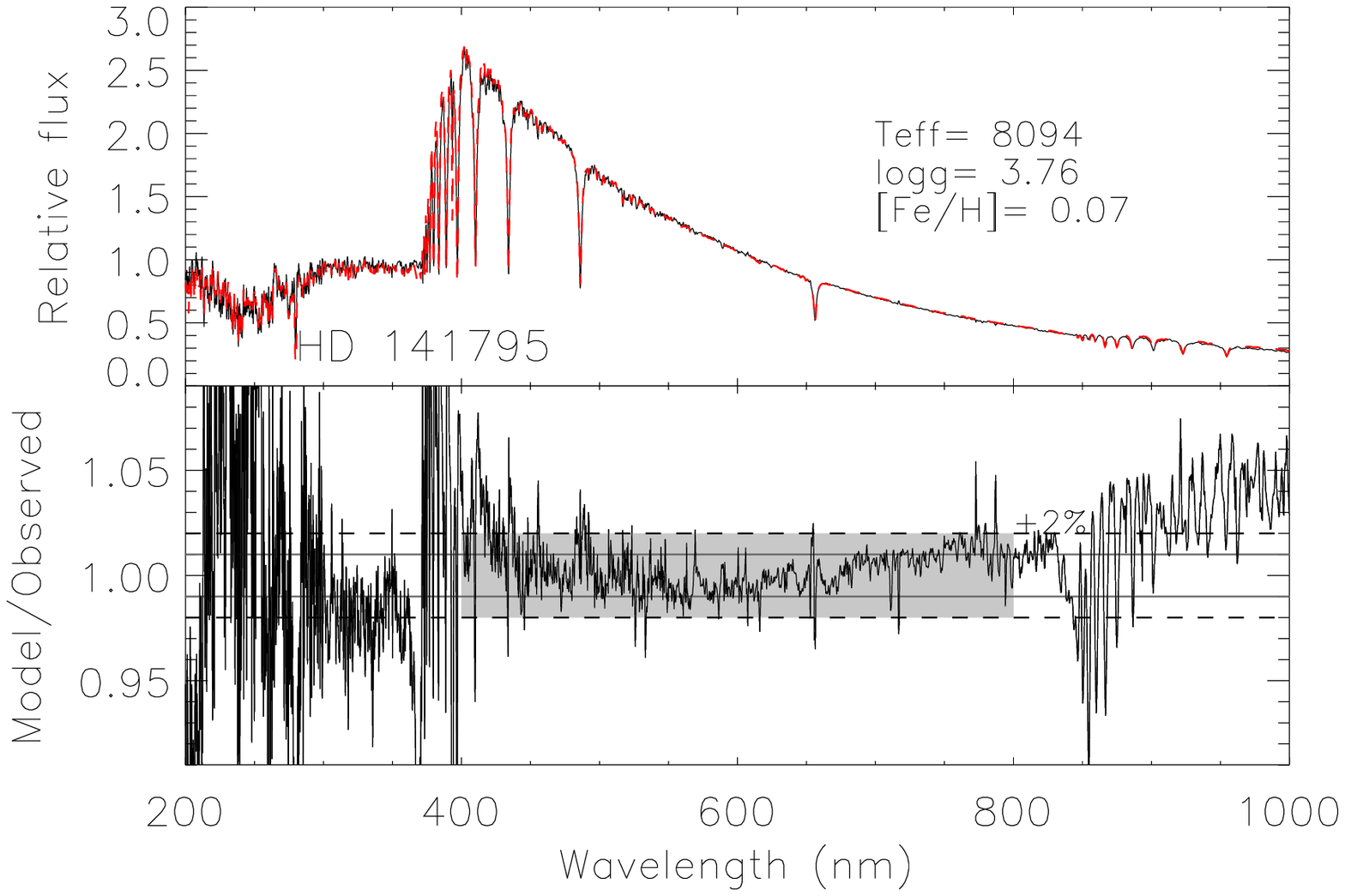}
 \caption{Upper panel: Observed (solid black line) and best-fitting model (broken red line) for 
 four representative stars in our sample of proposed flux standards. Bottom panel: 
 Ratio between model and observed fluxes.}
\label{f2}
\end{figure*}

The analysis of the NGSL A-type stars was carried out exactly 
in the same way as  Vega's. We obtained parameters for all
of them and, in general, acceptable fittings to the observed
spectral energy distributions.
The analysis was performed both on the first (v1) 
and the second (v2) versions of the NGSL. The differences
in the quality of the fittings were modest, so we decided
to adopt the results based on the updated version (v2).

We find values for the resolving power of the spectra in the library 
between 460 and 730, but with a narrow distribution, centered
at $R=534$, with a standard deviation of 60, and therefore we decided to 
force $R=500$ in the final fittings. We compared the results for the parameters
determined with the individual $R$ values for each star and $R=500$
and found negligible differences. We also examined spectra with higher resolution 
from the MILES library (Falc\'on-Barroso et al. 2011 and references therein) for
stars in common, arriving at a similar value for the resolving power ($R=597\pm 72$ K).
We selected stars with an rms scatter 
between model and observed spectra in the range 400-800 nm lower than 2\%, 
and a maximum excursion from unity in the same range within $\sim$3\%, 
derived parameters within a safe distance of the grid edges, 
and estimated distances under 200 pc, retaining 20 stars. 

We verified that integrating over the Johnson $V$ bandpass (Cohen et al. 2003) the fluxes
for the NGSL were consistent with published photometry in the Mermilliod
et al. (1997) catalog, identifying a significant discrepancy ($\sim 0.2$ mag) 
for HD 193281, which we discarded after finding an even larger spread in 
the literature for this star ($6.3<V<6.9$ mag). We also checked for 
variability in the Hipparcos Catalog
(Perryman et al. 1997), discarding one additional star (HD 112413)
with an rms scatter in $H_p$ larger than 0.06 mag. 
Only these 18 stars are discussed in the remainder of the paper.

Figure 2 shows representative fittings for 4 stars in the sample. 
Table 1 provides our derived atmospheric parameters for the 
complete sample, as well as those listed in the headers of the
FITS files of the NGSL. The agreement between the effective temperatures 
and metallicities is fair, with our temperatures being an average of 161 K cooler 
($\sigma=181$ K), and our metallicities an average of 0.12 dex lower ($\sigma=0.15$ dex).
The surface gravities show a similar offset, ours being 
lower by 0.18 dex, and an rms scatter $\sigma=0.16$ dex, 
but the discrepancies seem to be largest (by up to 0.5 dex) 
for the coolest and more metal rich stars in the sample. 

Our conclusion about the systematics in our derived surface gravities is 
reinforced by comparing
our results with Allende Prieto \& Lambert (1999), who compared $V$-band absolute 
magnitudes, computed from Hipparcos' parallaxes, and $B-V$ colors, with models of 
stellar interiors. Their surface gravities for 8 stars in common 
are an average of 0.33 larger than ours ($\sigma=0.18$ dex), and show a similar
trend as we found with the NGSL values, with larger discrepancies for cooler
more metal-rich stars. 

The comparison with metallicities  from the PASTEL
bibliographical catalog (Soubiran et al. 2010), which compiles  
stellar atmospheric parameters determined from high resolution,
high signal-to-noise spectra in the literature, shows, for eight stars in common, 
a mean offset of $0.02$ dex, ours being lower, and significant scatter ($\sigma=0.48$ dex).
Our gravities are also lower than those in PASTEL by an average of 0.08 dex
($\sigma=0.21$ dex). Lastly, our comparison with the parameters published
by Koleva \& Vazdekis (2012) shows the largest spread, 
with standard deviations of 702 K in $T_{\rm eff}$, 0.5 dex in $\log g$, 
and 0.6 dex in metallicity.

To evaluate the effect of neglecting reddening in our analysis, we have
tested the consequences of adopting instead the reddening estimates provided
by Lindler \& Heap (2008) as part of the NGSL, correcting the spectra for
reddening using the mean Galactic extinction curve given by Fitzpatrick (1999),
and fitting again the spectra with models. On average, the newly derived
effective temperatures where $+205 K$ ($\sigma=170$ K) warmer, the surface
gravities 0.10 dex ($\sigma=0.09$) higher, and the metallicities 0.08 ($\sigma=0.11$)
higher than the original values.  These offsets are comparable  
to the scatter, and consistent with our estimates of the systematic uncertainties. 
We therefore conclude that neglecting reddening is not a significant source of
error.

We stress that even though we derive atmospheric parameters as part of
our fitting process, our surface gravities and metallicities depend heavily 
on the Balmer and Paschen jumps and the near-UV line absorption, respectively,
so they should not be taken as indicative of the true parameters of the
stars, but merely as the values that provide the closest match to the
NGSL spectral energy distributions. We only use them
to select stars that are vetted as flux standards in the 400-800 nm window,
and to set their reference fluxes, after scaling the models to match the observations
in that window. We also use the theoretical fluxes to predict angular diameters from
the comparison with the observed ones, as described in \S \ref{diameter}, and 
although we find good agreement for the few stars with interferometric measurements,
care must be exercised when adopting these results for the stars for which 
the surface gravities we derive spectroscopically and those from stellar structure
models (the NGSL values in Table 1) differ. 
HD 18769 shows the highest difference in surface gravity, our value being lower than
that from stellar structure models listed in the NGSL. 
Other cases for which our values are significantly smaller are 
HD 319, HD 38237, HD 141795, and HD 201377.

\section{Absolute fluxes}
\label{absolute}

The absolute fluxes for the spectra in the NGSL are on the HST scale, but
some caveats apply, namely E1 aperture corrections, fringing, and 
red light contamination to UV (G230LB) fluxes. The authors of the library
have gone beyond the standard pipeline reduction to minimize the
impact of these problems, but it is possible that the systematic errors 
involved 
are significantly larger than for other STIS observations optimized 
for spectrophotometric calibration (see Bohlin \& Proffitt 2015). 
We have computed the magnitudes 
in the Johnson $V$ band using the response by
Cohen et al. (2003). These are provided in Table 1, together with the 
Johnson $V$-band magnitudes from Mermilliod et al. (1997). 

Vega is the nominal zero point for the Johnson and other photometric
systems, but as the best determination of the Vega spectrum
evolves with time, so does the Johnson Vega magnitude and zero 
point -- the photometry in that system remains relative to Vega,
but changing Vega's magnitude is a better option than changing the
magnitudes of all other stars 
(see the discussion by Ma\'{\i}z-Apellaniz 2006, 2007). 

For Vega's STIS spectrum of Bohlin \& Gilliland (with the corrections
described by Bohlin 2007) we find a $V=0.0226$ mag
on the Landolt scale (using the response and in-band flux 
provided by Cohen et al. 2003), in agreement with Bohlin's calculation 
($V=0.023 \pm 0.008$)\footnote{We note that a recent photometric study 
by Bohlin \& Landolt (2015) shows evidence that the $V$-band response
curve  published by Bessell \& Murphy (2012), shifted by $-20$ \AA, gives more
consistent results for about a dozen standard stars, and leads to a $V$ magnitude
for Vega of 0.028.}. 
We have similarly calculated the magnitudes
of the NGSL (v2) spectra, finding good agreement between them and 
the $V$ magnitudes collected from Mermilliod et al. (1997):
$< V - V_{\rm NGSL} > =   +0.008   \pm 
0.018$ mag\footnote{The quoted uncertainty is the standard deviation}.

This comparison suggests that the zero-point of the NGSL scale is correct
for our sample. Therefore, we propose to promote the stars in Tables 1 
to flux standards in the 400-800 nm spectral window. Their best-fitting models
in that spectral window are available online 
with this paper and from our website\footnote{http://hebe.as.utexas.edu/std/}.

\section{Predicted angular diameters}
\label{diameter}

The ratio between the  NGSL  fluxes scaled with the photometry and
the computed fluxes at the surface of the star in the 400-800 nm
window can be used to calculate the stellar angular diameters, 
exactly in the same way as described for Vega in \S \ref{vega}.

The results are given in Table 1. In two cases, these results
can be directly compared with angular diameters from
 the Palomar Testbed Interferometer (PTI) compiled 
by van Belle et al. (2008). 
These are for HD 97633 and HD 141851, for which 
we have derived angular diameters of $0.757  \pm  0.005$ 
and $0.398 \pm 0.003$ mas, respectively, while
the interferometric values are $0.784 \pm 0.037$ 
and $0.427 \pm 0.015$ mas, respectively. Maestro et al. (2013) have
recently measured the angular diameter for HD 97633 using 
CHARA, finding $0.740 \pm 0.024$ mas, so our value is bracketed by the 
PTI and CHARA measurements.
Boyajian et al. (2012) has also published an angular diameter
of $0.768 \pm 0.017$ mas for HD 141795, for which we find
$0.756 \pm 0.007$ mas.
The agreement between the two methods is good and
within uncertainties, but our estimated uncertainties are significantly 
smaller than those for the interferometric measurements.

There are two more objects with angular diameters in the van Belle et al.
list which have not made our quality cut described in \S 3.2: 
HD 28978  with $0.296 \pm 0.013$ mas, for which we would derive an angular
diameter of $0.272 \pm 0.005$ mas, and HD 163641, with $0.178 \pm 0.027$ mas, 
for which we would obtain $0.174 \pm 0.006$ mas. Again, we find good 
consistency within error bars, but our value for HD 28978 is somewhat smaller
than the interferometric diameter.

As discussed in \S \ref{stars}, there are a few stars for which 
the surface gravities we derived spectroscopically are significantly lower
from the NGSL values derived by Lindler \& Heap (2008) from the comparison with stellar
evolution models. As a check, we adopted the NGSL values, refit the spectra
with only $T_{\rm eff}$ and [Fe/H] as free parameters, and rederived angular diameters.
We find that for the stars for which our gravities were up to 0.3 dex different, the
inferred angular diameters differed only by less than 0.006 mas from those in 
Table 1. 


The NGSL gravities, including for HD 141795, are in excellent agreement with 
the values reported by Allende Prieto \& Lambert (1999) using a similar technique,
comparing absolute magnitudes from Hipparcos parallaxes with stellar models.
The few discrepancies between these {\it evolutionary} gravities and those we derive from
the spectrum may be  associated with fast rotation or other modeling
issues mentioned in the introduction.

\section{Summary and conclusions}
\label{summary}

We carefully examine the A-type stars with effective temperatures between
7,500 and 12,000 K in the Next Generation Stellar Library (Gregg et al. 2006; 
Heap \& Lindler 2007; Lindler \& Heap 2008) in the light of model atmospheres for appropriate
atmospheric parameters. We identify eighteen stars that we propose as 
flux calibrators for which the models 
closely reproduce the observations at the 3\% level in the 400-800 nm region.
These stars have $V$ magnitudes in the range 3 to 8, are spread in the sky ($-40< \delta < +27$ deg), 
and can be useful when absolute flux standards are required to calibrate spectroscopic 
observations that demand extremely high signal-to-noise ratios and  high dispersion.

The zero-point flux scale of A-type stars of the NGSL in the $V$ band appears
to be consistent with the HST scale (see Bohlin 2010). 
Despite departures from LTE and fast rotation, 
classical model atmospheres can typically match the spectral energy distribution of A-type
stars in the 400-800 nm range to better than 3 \%. This conclusion encourages
their adoption as primary flux standards for instruments targeting very bright
sources, such as observations of nearby stars hosting exoplanets.

Accurate stellar radii are necessary to derive
accurate planetary radii from (spectro-)photometric transits (del Burgo et al. 2010).
Our study underlines the potential to derive stellar radii from spectrophotometric
observations combined with accurate trigonometric parallaxes, such as those 
soon to be provided by Gaia.

\section{acknowledgements}

The authors are grateful to the referee, Ralph Bohlin, for comments that
helped to improve the paper.
CAP is thankful to the Spanish MINECO for support through grant
AYA2014-56359-P. This work has been supported by Mexican CONACyT research grant CB-2012-183007.
This research has made use of the SIMBAD database,
operated at CDS, Strasbourg, France and NASA's Astrophysics Data System.

\label{lastpage}

\end{document}